\documentclass[letterpaper,twocolumn,10pt]{article}
\usepackage{usenix2019_v3}

\usepackage{silence}
\PassOptionsToPackage{hyphens}{url}
\usepackage[hyphens]{url}
\PassOptionsToPackage{breaklinks,colorlinks}{hyperref}
\usepackage[breaklinks,colorlinks]{hyperref}
\PassOptionsToPackage{usenames,dvipsnames}{xcolor}
\usepackage[usenames,dvipsnames]{xcolor}
\usepackage{epigraph}
\hypersetup{
  colorlinks,
  linkcolor={red!50!black},
  citecolor={blue!50!black},
  urlcolor={blue!50!black}
}
\PassOptionsToPackage{usenames,dvipsnames}{xcolor}
\usepackage[usenames,dvipsnames]{xcolor}
\usepackage{colortbl}
\usepackage{amsmath,amsopn,amssymb,amsthm}
\usepackage{subfig}
\usepackage{endnotes,microtype,xspace,graphicx,fancyvrb,multirow}
\usepackage{booktabs}
\usepackage{array,underscore,relsize}
\usepackage[T1]{fontenc}
\usepackage{fancyhdr}
\usepackage{enumitem}
\usepackage{csquotes}
\usepackage[labelfont=bf,font=small]{caption}
\usepackage[normalem]{ulem}
\usepackage{tikz}
\usepackage{mathdots}
\usepackage{cancel}
\usepackage{color}
\usepackage{tabularx}
\usetikzlibrary{fadings}
\usepackage{calligra}

\usepackage{fp}

\usepackage[normalem]{ulem}
\usepackage{tcolorbox}

\usepackage{balance}

\usepackage{pifont}
\usepackage{wasysym}
\usepackage{cite}
\usepackage{algorithmic}
\usepackage{textcomp}
\usepackage{bm}
\usepackage{xfrac}
\usepackage{wrapfig}
\usepackage{multicol}

\usepackage{todonotes}

\usepackage{makecell}

\newcommand*{\origrightarrow}{}

\renewcommand*{\textrightarrow}{\fontfamily{cmr}\selectfont\origrightarrow}

\newcommand{\cc}[1]{\mbox{\smaller[0.5]\texttt{#1}}}

\usepackage{etoolbox}
\let\oldbibitem\bibitem
\def\bibitem{\vfill\oldbibitem}

\fvset{fontsize=\scriptsize,xleftmargin=8pt,numbers=left,numbersep=5pt}

\input{code/fmt}

\def\Snospace~{\S{}}

\newcommand{\x}{$\times$\xspace}

\newif\ifdraft\drafttrue
\newif\ifnotes\notestrue
\ifdraft\else\notesfalse\fi

\input{glyphtounicode}
\newcolumntype{R}[1]{>{\raggedleft\let\newline\\\arraybackslash\hspace{0pt}}p{#1}}

\newcommand{\includepdf}[1]{
  \includegraphics[width=\columnwidth]{#1}
}

\newcommand{\squishlist}{
\begin{itemize}[noitemsep,nolistsep]
  \setlength{\itemsep}{-0pt}
}
\newcommand{\squishend}{
  \end{itemize}
}

\newcommand{\squishlists}{
\begin{itemize}[leftmargin=3mm,noitemsep,nolistsep]
  \setlength{\itemsep}{-0pt}
}
\newcommand{\squishends}{
  \end{itemize}
}

\usepackage{tikz}

\usepackage{xstring}
\newcommand{\PP}[1]{
\vspace{2px}
\noindent{\bf \IfEndWith{#1}{.}{#1}{#1.}}
}

\usepackage{xstring}

\newcommand{\etal}{\textit{et al}.\xspace}
\newcommand{\ie}{\textit{i}.\textit{e}.}
\newcommand{\eg}{\textit{e}.\textit{g}.}

\newcommand{\boxbeg}{
\noindent
\begin{tabular}{|l|}\hline
\begin{minipage}{0.94\columnwidth}
\vspace{3px}
\noindent
}

\newcommand{\boxend}{
\vspace{2px}
\end{minipage}\\ \hline
\end{tabular}
\vspace{0.5px}
}

\newtheorem*{theorem1*}{Theorem}

\newtheorem*{lemma1*}{Lemma}

\newcommand{\clwb}{\mbox{\cc{clwb}}\xspace}
\newcommand{\sfence}{\mbox{\cc{sfence}}\xspace}

\newcommand{\yesmark}{\checkmark}
\newcommand{\nomark}{{\small $\times$}\xspace}

\newcommand{\load}{\mbox{\cc{load}}\xspace}
\newcommand{\store}{\mbox{\cc{store}}\xspace}

\newcommand{\flush}{\mbox{\cc{flush}}\xspace}
\newcommand{\fence}{\mbox{\cc{fence}}\xspace}

\newcommand{\hb}{\mbox{$\xrightarrow[]{\text{hb}}$}\xspace} 
\newcommand{\cd}{\mbox{$\xrightarrow[]{\text{cd}}$}\xspace} 
\newcommand{\dd}{\mbox{$\xrightarrow[]{\text{dd}}$}\xspace} 

\definecolor{darkolivegreen}{rgb}{0.33, 0.42, 0.18}

\newcommand{\cppX}[1]{\cc{{\color{darkolivegreen}\textbf{#1}}}}
\newcommand{\UPST}{\mbox{$\mathbb{U}$}\xspace} 
\newcommand{\PST}{\mbox{$\mathbb{P}$}\xspace}  

\newcommand{\lhinsertkey}{14}
\newcommand{\lhinsertval}{15}
\newcommand{\lhinserttoken}{18}
\newcommand{\lhinsertpflushk}{20}

\newcommand{\lhinsertfencekv}{23}

\newcommand{\lhupdatekey}{34}
\newcommand{\lhupdateval}{35}
\newcommand{\lhupdatetokenj}{38}
\newcommand{\lhupdatetokenk}{39}

\newcommand{\lhquerytoken}{3}

\newcommand{\lhqueryval}{7}

\newcommand{\sys}{\mbox{\textsc{Witcher}}\xspace}

\newcommand{\NumOfOrderBugs}{19\xspace}
\newcommand{\NumOfAtomicityBugs}{18\xspace}
\newcommand{\NumOfTotalBugs}{37\xspace}
\newcommand{\NumOfNewBugs}{32\xspace}

\newcommand{\NumOfAppsTested}{17\xspace}
\newcommand{\NumOfAppsBugDetected}{13\xspace}

\gdef\thedate{2020-12-10 19:41:38 -0500}

\begin{document}
\title{
\sys: Detecting Crash Consistency Bugs in Non-volatile Memory Programs
}

\newcommand{\sbu}{{\textsuperscript{$\dagger$}}}

\author{
    Xinwei Fu \quad
    Wook-Hee Kim \quad
    Ajay Paddayuru Shreepathi\sbu \\
    Mohannad Ismail \quad
    Sunny Wadkar \quad
    Changwoo Min \quad
    Dongyoon Lee\sbu \\
    Virginia Tech \quad \quad
    {\sbu}Stony Brook University
}



\date{}
\maketitle
\thispagestyle{empty}

\begin{abstract}
The advent of non-volatile main memory (NVM) enables the development
of crash-consistent software without paying storage stack
overhead. However, building a correct crash-consistent program remains
very challenging in the presence of a volatile cache.
This paper presents \sys, a crash consistency bug detector for NVM
software, that is (1) scalable -- does not suffer from test space
explosion, (2) automatic -- does not require manual source code
annotations, and (3) precise -- does not produce false positives.
\sys first infers a set of ``likely invariants'' that are
believed to be true to be crash-consistent by analyzing source codes
and NVM access traces. \sys automatically
composes NVM images that simulate those potentially inconsistent
(crashing) states violating the likely invariants.
Then \sys performs ``output equivalence checking'' by comparing the
output of program executions with and without a simulated crash.
It validates if a likely invariant violation under test is a true crash
consistency bug.
Evaluation with ten persistent data structures, two real-world
servers, and five example codes in Intel's PMDK library shows
that \sys outperforms state-of-the-art tools. \sys discovers
\NumOfTotalBugs (\NumOfNewBugs new)
crash consistency bugs, which were all confirmed.
\end{abstract}

\section{Introduction}
\label{s:intro}

Non-volatile main memory (NVM) technologies, such as the recently
commercialized Intel Optane DC Persistent
Memory~\cite{optanedc-announce-web, 3dxpoint-micron}, provide
persistence of storage along with traditional DRAM characteristics
such as byte addressability and low access latency. NVMs are attached
to processors via a memory bus so that programs can access the NVMs
using regular \load and \store instructions. The ability to directly
access NVMs provides a new opportunity to build
\emph{crash-consistent} software without paying storage stack
overhead. Programs can recover a consistent state from a
potentially-inconsistent persistent NVM state in the event of an
application or a system crash, or a sudden power loss (hereafter crash
for brevity).

However, designing and implementing a correct crash-consistent
program is challenging. NVM data on a processor's volatile cache
may not be persisted after a crash. This implies that the completion
of a store instruction does not guarantee the persistence of
memory. Furthermore, a processor cache can evict cache lines in an
arbitrary order so that the NVM memory locations may not be persisted
in the same order as the program (store) order. Lastly, the current
ISA does not provide an atomic instruction to update multiple NVM
locations.

Therefore, ensuring crash consistency requires a developer to
explicitly add a cache line flush and store fence instructions (\eg,
\clwb and \sfence in x86 architecture) and to devise a custom
mechanism to ensure ordering and atomicity guarantee, making NVM
programming hard and error-prone.

Recently, several solutions for detecting crash consistency bugs have been
proposed, but they are not satisfactory.
%
One line of tools~\cite{yat-lantz-atc14, pmreorder-intel-web,
scmtest-oukid-damon16} attempts to test all possible inconsistent
states exhaustively, leading to a scalability issue. For instance,
Yat~\cite{yat-lantz-atc14} reports that a program execution with 14K
cache line flushes leads to 789 million combinations to test, which
will take roughly 5.2 years.
%
Another family~\cite{pmtest-liu-asplos19, xfdetector-liu-asplos20}
asks developers to annotate correctness conditions
manually. Annotating a large NVM program is challenging, potentially
resulting in both false negatives (missing annotation) and false positives
(incorrect annotation) depending on the quality of annotations.

This paper presents \sys, a new crash consistency bug detector
for NVM software. \sys is (1) scalable -- it does not suffer from test
space explosion, (2) automatic -- it does not require manual source
code annotations, and (3) precise -- it does not produce false
positives. We build \sys based on the following two key insights.


First, \sys automatically infers a set of \emph{likely program invariants} that
are believed to be true to be crash-consistent by analyzing source codes and
execution traces, with a hypothesis that programmers leave some hints on what
they want to ensure.  \sys then tests only for those likely invariants, instead
of relying on exhaustive testing or user's manual annotation. 
For example, for NVM addresses \cc{X} and \cc{Y}, from the source code
where the write of \cc{Y} is control-dependent on the read of \cc{X} 
(\eg, \cc{\cppX{if}(X)\{Y=3\}}), \sys infers a likely invariant that ``\cc{X}
should be persisted before \cc{Y}'', assuming that developers would not want to
update persistent data \cc{Y} based on unpersisted data \cc{X}. Then it
checks the persistence ordering between \cc{X} and \cc{Y}, but not between
\cc{X} and another irrelevant \cc{Z}, saving the testing time.

Second, we also automatically validate those likely invariants (\ie,
to check if a violation of those invariants is a true crash
consistency bug) by performing \emph{output equivalence checking}.
If a program is crash-consistent, it should produce the same output
(\eg, \cc{query(k)}) between two executions with and without a
crash. Based on this observation, \sys composes a set of crash NVM
images, each of which simulates a potentially inconsistent (crashing)
state violating a likely invariant. Then it runs a randomly generated
test case (\eg, a sequence of mixed \cc{insert}, \cc{delete}, and
\cc{query}), and compares the output of program executions with and
without a simulated crash. Any mismatch (\eg, one leads to a fault, or
produces a different output) is a definite clue of a
bug.


\sys considers two forms of likely invariants ensuring (1) persistence ordering
(\eg, \cc{X} should be persisted before \cc{Y}) and (2) persistence atomicity
(\eg, \cc{X} and \cc{Y} should be persisted atomically). Then it tests if
persistence primitives (\eg, cache line flush and store fence
instructions) are used properly to ensure them. 
This approach allows \sys to detect crash consistency bugs in both
\emph{low-level} NVM programs, which use the assembly-level persistence primitives, and
\emph{transactional} NVM programs, which rely on a logging logic in a
transaction library (\eg, Intel's Persistent Memory Development Kit
(PMDK)~\cite{pmdk}), in a unified manner. 
Under the hood, such transactional libraries use the same flush and fence
primitives for persistence.
\sys traces and analyzes PMDK internals such as persistence heap allocation and
transactional undo logging logics, validating both the PMDK library itself and
transactional NVM programs using them.
Also, \sys supports both single-threaded and multi-threaded testing.

We evaluated \sys with \NumOfAppsTested NVM programs consisted of 92K lines of code
(LOC), which include ten highly-optimized persistent data structures
(appeared in top-tier systems conferences), two server applications
Redis~\cite{pmem-redis-git} and Memcached~\cite{pmem-memcached-git}
that are ported with PMDK, and five example code included in PMDK.
Using randomly generated test cases, \sys detected \NumOfTotalBugs (\NumOfNewBugs new) crash
consistency bugs in \NumOfAppsBugDetected programs, all of which are confirmed by the
developers. 
One new bug was found in the PMDK's persistent pool/heap management library.
%

We make the following contributions in this paper:
\squishlists

\item We propose a \emph{likely invariant-based approach} to infer
  likely correct conditions to detect crash consistency bugs without
  manual annotations or exhaustive testing.

\item We present an \emph{output equivalence-based technique} to
  identify an incorrect execution without user-provided consistency
  checkers or annotations.

\item We implement \sys that detects crash consistency bugs in a
  scalable, automatic, and precise manner using likely invariant inference and
  output equivalence checking.

\item Our evaluation shows that \sys detects
  \NumOfTotalBugs (\NumOfNewBugs new) confirmed bugs
  and outperforms existing solutions: 
  it discovered more bugs in a scalable and automatic manner.
  All the bugs reported were true positives.

\squishends

\section{Motivation}
\label{s:motiv}

This section first demonstrates the types of crash consistency bugs
along with real-world examples that \sys found in an NVM-optimized
resizable hash table, Level
Hashing~\cite{level-hashing-zuo-osdi18}. Then, we discuss the
limitations of existing testing techniques.

\subsection{Crash Consistency Bugs}
\label{s:motiv:examples}

\begin{figure}[t!]
  \centering
  \includegraphics[width=.96\columnwidth]{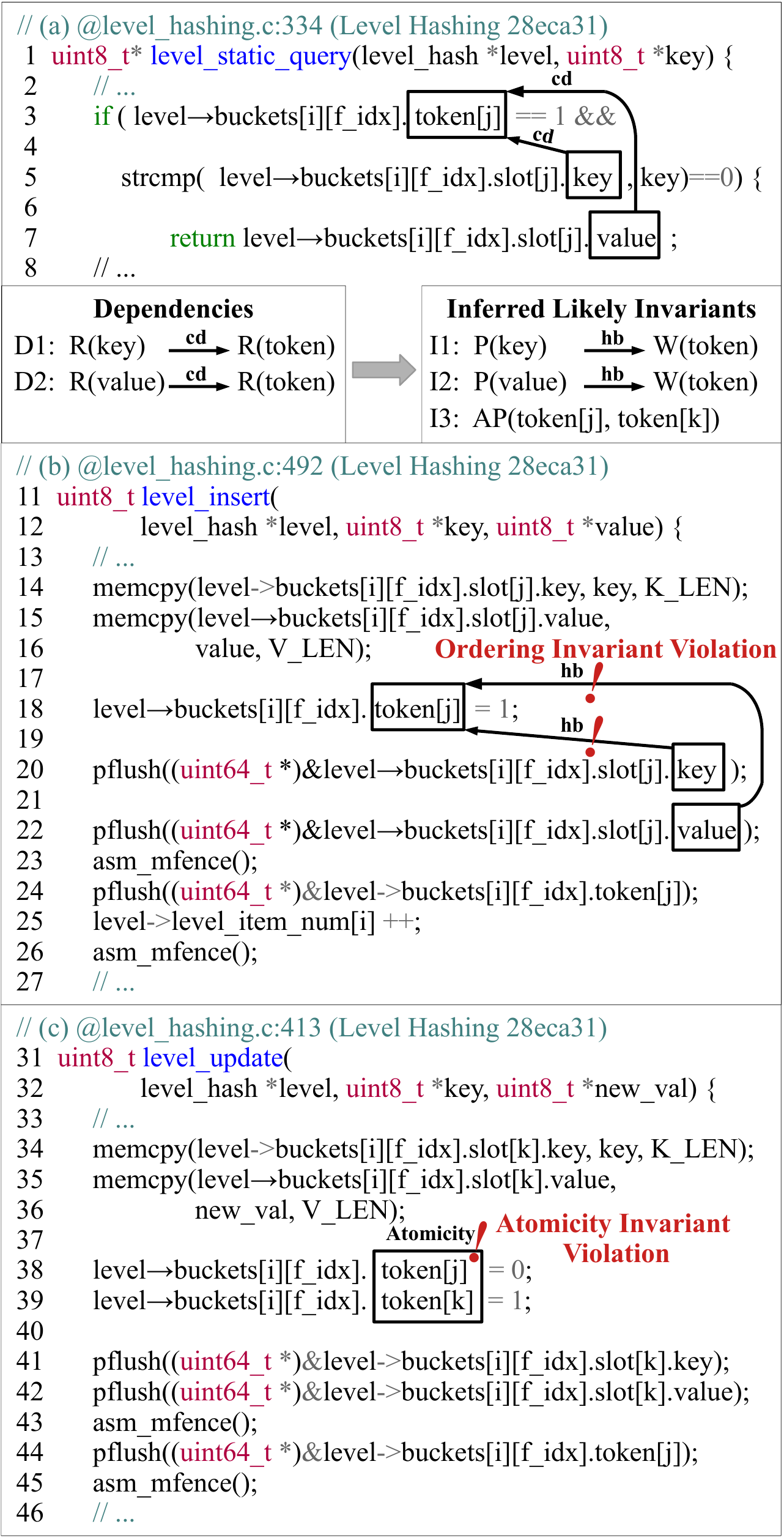}
  \vspace{-2px}
  \caption{
    Using the likely invariants inferred from (a), \sys detects two
    crash consistency bugs (b) and (c) in Level
    Hashing.
  }
  \vspace{-8px}
  \label{f:lh-bugs}
\end{figure}


\PP{(1) Persistence ordering violations}
A buggy NVM program may not maintain proper \emph{persistence
ordering} when updating multiple NVM locations. A processor cache can
evict cache lines in an arbitrary order, and a processor can re-order a
cache line flush instruction. As a result, the program order among
multiple {\store}s may mismatch with the persistence order. If a
developer wants to ensure that one \store becomes persisted before
another, she has to explicitly add a cache line flush followed by a
store fence instruction (\eg, \clwb and \sfence in x86 architecture)
between them. 
We consider a missing persistence primitive bug (used in
XFDetector~\cite{xfdetector-liu-asplos20} and RECIPE~\cite{recipe-lee-sosp19})
as a kind of persistence ordering bugs because it voids an ordering guarantee.

We found ten persistence ordering bugs (\autoref{s:eval}) in Level
Hashing~\cite{level-hashing-zuo-osdi18}. For performance, 
Level Hashing introduces \emph{log-free} write
operations. It maintains a flag token for each key-value slot where
token 0 denotes the corresponding key-value slot is empty and token 1
denotes non-empty.
\autoref{f:lh-bugs}(b) shows the
\cc{level_insert} function. It intends to update the key-value slot
(Lines~\lhinsertkey,~\lhinsertval) before
updating token
(Line~\lhinserttoken).
However, if a crash happens after the token's cache line is evicted
(thus persisted) but before the key-value slot's cache line is not
(before Line~\lhinsertpflushk), an inconsistent state could occur -- the
token indicates that the corresponding key-value slot is non-empty, but
the slot is never written to NVM. Thus, the garbage value can be read
(as in \autoref{f:design-example}(h)), implying that the insert operation failed to
provide an atomic (all or nothing) semantic upon a crash. The persistent barrier at
Lines~\lhinsertpflushk-\lhinsertfencekv~should be moved before
updating the token at Line~\lhinserttoken.

\PP{(2) Persistence atomicity violations}
A buggy NVM program may not correctly enforce \emph{persistence
atomicity} among multiple NVM updates. If a program crashes in the
middle of a sequence of NVM updates, an inconsistent state may
occur. We found 16 atomicity bugs in seven NVM programs
(\autoref{s:eval}).

\autoref{f:lh-bugs}(c) shows a persistence atomicity bug found in
Level Hashing's \cc{level_update} function.
Level Hashing opportunistically performs a log-free update. If there is
an empty slot in the bucket storing the old key-value slot, a new slot
is stored to the empty slot (Lines~\lhupdatekey,~\lhupdateval), and
then the old and new tokens are modified
(Lines~\lhupdatetokenj,~\lhupdatetokenk). Since the new slot is not
overwritten to the old slot, Level Hashing can avoid the costly
logging operations.
However, the code incorrectly assumes that updating two tokens is
atomic. If a crash happens right after turning off the old token
(Line~\lhupdatetokenj) and the cache line of the old
token is evicted (persisted), the crash consistency problem happens.
Since the old token is persisted with 0 but the new token
(Line~\lhupdatetokenk) is not turned on, we
permanently lose the updating key.
To solve this bug, we have to persist two
tokens atomically. In this example, we can place two tokens in an
8-byte word
using bit representation
and update them with a single 8-byte store to update two tokens
atomically. 


\subsection{Limitations of Existing Solutions}
\label{s:motive:limitations}

\PP{(1) Huge testing space}
\emph{Exhaustive testing approach}~\cite{yat-lantz-atc14,
pmreorder-intel-web, scmtest-oukid-damon16} attempts to permute all
possible persistent states on a crash. Then, for each crashed state,
they rely on a user-provided consistency checker to validate whether
NVM data is consistent.
%
%
%
However, they often do not scale. For instance, the testing space of
Yat~\cite{yat-lantz-atc14} explodes exponentially in the number of
\store instructions. For example, testing Level Hashing with 2000 (random) operations, the
Yat tests $10^{77}$ total permutations (see the
detailed discussion in~\autoref{s:eval:ts}).
%

\PP{(2) Manual annotation burden}
The test space explosion problem motivated the recent
\emph{annotation-based approach}, such as
PMTest~\cite{pmtest-liu-asplos19} and
XFDetector~\cite{xfdetector-liu-asplos20}.
%
%
Although these approaches are fast without exhaustive testing, it puts
a significant annotation burden on the developers, raising soundness,
completeness, and scalability concerns. A missing annotation may miss
crash consistency bugs (false negatives). The wrong annotation may
produce false bugs (false positives). Annotating a large NVM software
soundly and precisely is very challenging. Lastly, we note that PMTest
lacks support for detecting persistence atomicity violations such as
\autoref{f:lh-bugs}(c), and XFDetector cannot support low-level NVM programs
that do not rely on logging such as Level Hashing.

\begin{figure*}[t!]
  \centering
  \includegraphics[width=\textwidth]{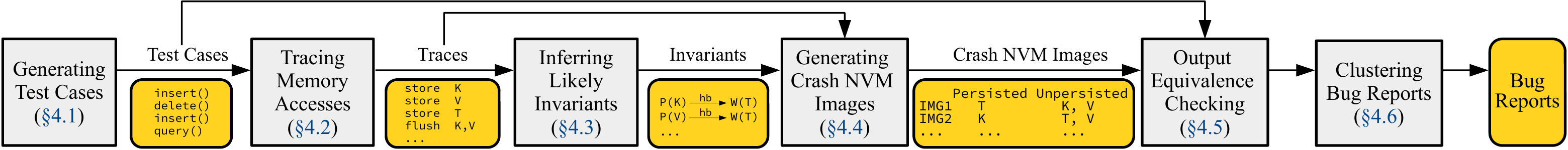}
  \vspace{-15px}
  \caption{
    The architecture of \sys.
  }
  \vspace{-12px}
  \label{f:arch}
\end{figure*}

\PP{(3) Correctness validation after a crash}
Validating the correctness (crash consistency) after a crash is another
challenge.
%
%
The first approach relies on a user-provided consistency checker (\eg,
Yat~\cite{yat-lantz-atc14}, PMReorder~\cite{pmreorder-intel-web}).
However, the correctness
of a manually-written checker is often a concern; a recent
study~\cite{fs-rel-jaffer-atc19} reports that even checkers
(\cc{fsck}) of the production-grade file systems (ext4, F2FS, and
btrfs) cannot recover 16\% of failures.
The second approach relies on user-annotated invariant checkers (\eg,
PMTest~\cite{pmtest-liu-asplos19}). As discussed earlier, the soundness,
completeness, and scalability of user annotations is a
concern.
The last relies on the user's manual investigation without
providing automatic validation (\eg,
XFDetector~\cite{xfdetector-liu-asplos20}).
That would waste a lot of
developer’s time with a high false-positive rate.

\section{Our Approach}
\label{s:overview}


\subsection{Inference of Likely Invariants}
\label{s:overview:invariant}

We propose a novel \emph{likely invariant-based} approach to detect
NVM crash consistency bugs scalably and automatically. Our key
observation is that programmers often left some hints on what they
want to ensure in the source codes. Thus we can infer a
set of likely-correctness conditions, which we refer to as
\emph{likely invariants} (hereafter invariants for short), by
analyzing source codes and execution traces.

Using the Level Hashing example as mentioned earlier, let us
demonstrate how we can infer an invariant from the query (or lookup)
function \cc{level\_static\_query} in \autoref{f:lh-bugs}(a), and
apply it to find the crash consistency bugs in \cc{level\_insert} and
\cc{level\_update} in \autoref{f:lh-bugs}(b) and (c).
\cc{level\_static\_query} reads the key/value only if the token is
non-empty. In other words, there is a control dependency between the
read of a token and a key-value pair
(Lines~\lhquerytoken-\lhqueryval): \eg, we denote it as
\cc{R(slot[j].key){\cd}R(token[j])}. We analyze
the implication of this control dependency as follows.

We first refer to the common NVM programming pattern that uses
a flag (\cc{token}) to ensure the persistence atomicity of data
(\cc{key/value}) as \emph{guarded protection}.
We have observed this guarded protection pattern in many NVM programs
including key-value stores~\cite{pmemkv-web,hikv-xia-atc17,pmemcached-marathe-hotos17}, logging implementations~\cite{scalablelogging-wang-vldb14,nvramlog-huang-vldb14,wbl-arulraj-vldb16,justdolog-izraelevitz-asplos16,crls-huang-atc18},
persistent data structures~\cite{pb+tree-chen-vldb15,fptree-oukid-sigmod16,wort-lee-fast17,fast-fair-hwang-fast18,logfreeds-david-atc18,cceh-nam-fast19,recipe-lee-sosp19}, memory allocators~\cite{nvm-malloc-schwalb-vldb15,makalu-kumud-oopsla16,pmdk,pallocator-oukid-vldb17},
and file systems~\cite{btrfs-condit-sosp09,pmfs-dulloor-eurosys14,nova-xu-fast16,splitfs-kadekodi-sosp19,zofs-dong-sosp19}.
The guarded protection follows the following reader-writer pattern
around a flag variable, which we call \emph{``guardian''}: (1) The
writer ensures that both key and value are ``persisted before'' the flag
is persisted 
(\autoref{f:lh-bugs}(b)). (2) The reader checks if the flag is set
before reading the key and value, which we call \emph{``guarded
  read''} (\autoref{f:lh-bugs}(a)). The persistence ordering (for the
writer side) and the guarded read (for the reader side) together
ensure that the reader reads atomic (both old or both new) states of
key and value.

From the guarded read pattern in \autoref{f:lh-bugs}(a), we infer the first
\emph{persistence ordering invariant}: a key-value pair should be
persisted before a token (\ie,
\cc{P(slot[j].key/value){\hb}W(token[j])}). We then extend it 
to the second \emph{persistence atomicity invariant}: the updates of
two or more guardians should be atomic. Otherwise,
an atomic update of multiple key-value slots cannot be
guaranteed (\ie, \cc{AP(token[j],token[k])}).

Later we find that \cc{level\_insert} violates the persistence
ordering invariant at Line~\lhinserttoken, and \cc{level\_update}
violates the persistence atomicity invariant at Line~\lhupdatetokenk.
\sys tests only NVM states that violate the inferred
invariants. For example, in \cc{level\_insert} we test only one case that
a token is persisted but a key-value pair is not persisted,
which violates the writer pattern in the guarded
protection. Similarly, in \cc{level\_update} we test two cases that
one token is persisted and another token is not. In this way, we can
significantly reduce testing space without the developer's manual
annotation.

In \autoref{s:design:invariant}, we present more generalized meta-rules
to infer likely invariants beyond guarded protection. Note that \sys
does not require prior knowledge of truth and does not assume
invariants are always correct: if two invariants contradict, we test
both cases to discern which one is correct.

\subsection{Output Equivalence Checking}
\label{s:overview:outeq}

We propose an \emph{output equivalence checking} approach to validate
if an NVM state that violates an inferred invariant is indeed
inconsistent, indicating a crash consistency bug. The key
insight is that we can construct an oracle (a correct execution) to
compare with to discern whether a given NVM state is
inconsistent by leveraging atomic (all or nothing) semantics of
(correct) crash-consistent NVM programs. Hence we do not
need to rely on a user-provided consistency checker or a code
annotation for validation (as in existing works).

\begin{figure*}[t]
  \centering
  \includegraphics[width=\textwidth]{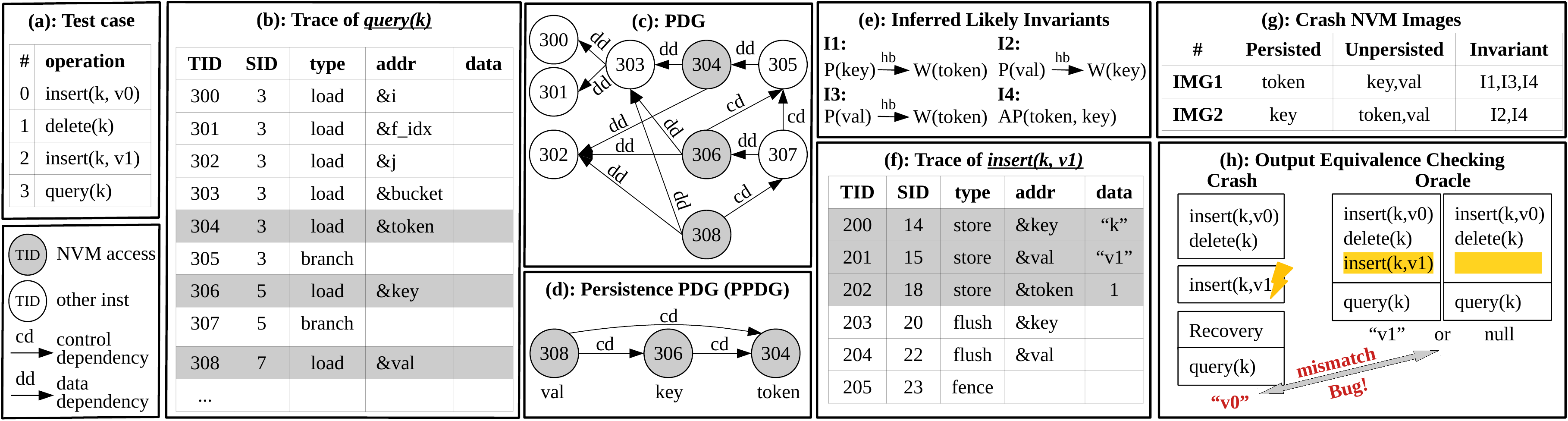}
  \vspace{-15px}
  \caption{An example of \sys's crash consistency bug detection steps.}
  \vspace{-12px}
  \label{f:design-example}
\end{figure*}

Suppose that we perform the following four operations on Level
Hashing: \cc{insert(k,v0)}, \cc{delete(k)}, \cc{insert(k,v1)}, and
\cc{query(k)} (see \autoref{f:design-example}(a)). If the program
crashes while executing the third \cc{insert(k,v1)}, it should behave as if the
operation either did happen or did not. After the resumption, we know
two possible, correct outputs of the following \cc{query(k)} are
either \cc{v1} or \cc{null}. If a program starting from an invariant-violating
NVM state produces output that is  
different from these two oracles, then we can confidently conclude
that the program is not crash-consistent, and the invariant violation is
indeed a true crash consistency bug.

Note that our oracles rely on test cases, and thus some crash
consistency bugs may not be detected if they do not produce visible
symptoms (\eg, segmentation fault, different output, etc.) on the
given test cases. This implies that we may have false negatives.
However, any detected output divergence is indeed an indicator of a
true crash consistency bug: \ie, we do not have false positives.

\section{Design of \sys}
\label{s:design}

We describe the detailed design of
\sys. \autoref{f:arch} illustrates \sys's six-stage architecture. As a
dynamic systematic testing tool, \sys starts by generating a random
test case (\autoref{s:design:tc}). \sys instruments an NVM program and
runs the test case to collect a memory trace
(\autoref{s:design:trace}), from which \sys infers invariants
(\autoref{s:design:invariant}). Then, from the same trace, \sys
constructs a set of legal crash NVM images violating the inferred
invariants (\autoref{s:design:ctp}) and performs output equivalence
checking to validate them (\autoref{s:design:validation}). Last, \sys
reports the crash consistency bugs after clustering similar cases
(\autoref{s:design:clustering}) to ease developers' root cause
analysis.
%
%
%
\sys supports both single-threaded and multi-threaded
testing. We explain single-threaded testing in this section. Then we
present our extension for multi-threading later
in~\autoref{s:design:multithread}.

\subsection{Generating Test Cases}
\label{s:design:tc}

\sys requires a deterministic test case such that its output is
deterministic for a given input for output equivalence
checking. Any deterministic test case with a good code coverage would
suffice.
An ideal crash consistency testing tool should explore the three
dimensions of testing space: (1) program input, (2) thread
interleaving, and (3) NVM state (persistence). \sys focuses on
exploring the (3) persistence space in a systematic and scalable
manner. 
We leave a smarter test case generation for program input and thread
interleaving and their test space reduction as future work, and instead use
random testing in this work.

By default, \sys supports random test case generation of index
structures, providing standard APIs such as \cc{insert}, \cc{delete},
\cc{update}, \cc{query}, and \cc{scan}. \sys randomly generates a list
of operations, keys, and values. For operation parameters, 
in order to make some dependent operations (\eg,
 \cc{get}, \cc{delete}, \cc{update}) more meaningful, we assign a higher probability
to generate a key used before in earlier operations. 

\subsection{Tracing Memory Accesses}
\label{s:design:trace}

\sys uses an NVM program execution trace to infer invariant
(\autoref{s:design:invariant}) and construct crash NVM images
(\autoref{s:design:ctp}) to validate later. We instrument an NVM
program using an LLVM compiler pass and then execute the instrumented
NVM binary with a test case to collect the execution trace.
%
We trace load, store (including the updated value), branch, call/return, flush (\ie,
\cc{clflush}, \cc{clflushopt}, and \cc{clwb}) and memory fence (\ie,
\cc{sfence} and \cc{mfence}) instructions.
%
%
%
We order the instructions in a trace by protecting our tracing
code using a global mutex so we can easily analyze traces of
multi-threaded programs.

Suppose we trace Level Hashing in~\autoref{f:lh-bugs} using the test
case with four operations in \autoref{f:design-example}(a).
~\autoref{f:design-example}(b)
shows the trace of the last \cc{last_static_query} operation. Each
trace includes a unique Trace ID (TID), a Static Instruction ID
(SID), which is essentially the instruction location in the binary,
and instruction type. For \load and
\store, \sys additionally traces its address, length (not shown), and
data (for \store), and whether accessing DRAM (white) or NVM (gray).

\subsection{Inferring Likely Invariants}
\label{s:design:invariant}

In this step, \sys performs a program dependency analysis and infers a
set of likely invariants from the memory trace.
We first describe a set of meta-rules designed to infer 
(1) persistence ordering invariants and (2) persistence atomicity
invariants (\autoref{s:design:invariant:logfree}).
Then we explain how
\sys uses program dependence analysis to apply the meta-rules
and infers the corresponding invariants from the memory trace
(\autoref{s:design:invariant:extraction}).

\subsubsection{Meta-Rules for Ordering and Atomicity Invariants}
\label{s:design:invariant:logfree}

At a high level, each rule looks for control and/or data
dependencies between NVM locations \cc{X} and \cc{Y}, 
and infers a likely invariant that ``\cc{X} should be
persisted before \cc{Y}'' (ordering invariant) or ``\cc{X} and \cc{Y}
should be persisted atomically'' (atomic invariant).
From the invariant, \sys later constructs an NVM state that violates it
-- \eg, ``\cc{Y} is persisted, but \cc{X} is not''
(\autoref{s:design:ctp}) and tests if this invariant violation
produces a wrong output (\autoref{s:design:validation}). 
Another way to explain the benefit of invariant inference is that for two
NVM addresses \cc{X} and \cc{Y}, if \sys does not detect any
dependency, it does not test such cases involving \cc{X} and \cc{Y}
and saves the test time (assuming that independent NVM objects do not
lead to an inconsistent state). \autoref{t:metarule} summarizes
meta-rules for inferring invariants:

\PP{RO1. Data dependency implies persistence ordering.}
Suppose a developer writes a code ``\cc{Y=X+3}'' where the write
of \cc{Y} is data-dependent on the read of \cc{X} (which we denote
\cc{W(Y)}{\dd}\cc{R(X)}). From the data dependency, we hypothesize that the
developer would want \cc{X} to be persisted before updating \cc{Y}
(\ie, \cc{P(X)}{\hb}\cc{W(Y)}) so that she does not update an NVM state
\cc{Y} based on ``unpersisted'' \cc{X}. Otherwise, an inconsistent
state may be generated when a program crashes after
\cc{Y} becomes persisted, but \cc{X} is not.
The first meta-rule \cc{RO1} in \autoref{t:metarule} is based on this
reasoning: for two memory locations \cc{X} and \cc{Y} in NVM, if we
find the Condition \cc{W(Y)}{\dd}\cc{R(X)}, we infer the Likely Invariant
\cc{P(X)}{\hb}\cc{W(Y)}. From the invariant we later create a crash NVM image
where \cc{Y} is persisted and
\cc{X} is unpersisted, if such state is legal in program execution.

\PP{RO2. Control dependency implies persistence ordering.}
Consider a code ``\cc{\cppX{if}(X) {Y=3}}'' where the write of
\cc{Y} is control-dependent on the read of \cc{X}.
Similar to \cc{RO1}, we hypothesize that a developer would
want to make \cc{X} persisted before updating \cc{Y}
so that she does not update \cc{Y} based on ``unpersisted''
\cc{X}.
More formally, \cc{RO2} says: for two memory locations \cc{X} and
\cc{Y} in NVM, if we find the Condition \cc{W(Y)}{\cd}\cc{R(X)}, we infer
the Likely Invariant \cc{P(X)}{\hb}\cc{W(Y)}. From the invariant, we
simulate a state where \cc{Y} is persisted but \cc{X} is
not if such state is legal. 

\PP{RO3. Guarded read implies persistence ordering.}
As discussed in~\autoref{s:overview:invariant}, guarded protection
is a common NVM programming pattern. It
achieves the atomicity of data using the writer-side persistence
ordering and the reader-side guarded read.
Based on this observation, if we see the reader-side guarded read
pattern, we infer the likely invariant on the writer-side
persistence ordering.
In other words, \cc{RO3} says: for two memory locations \cc{X} and
\cc{Y} in NVM, if we find the Condition \cc{R(Y)}{\cd}\cc{R(X)}, we infer
the Likely Invariant \cc{P(Y)}{\hb}\cc{W(X)}. We then validate the
invariant from the
NVM state such that \cc{X} is persisted but \cc{Y} is not.
Note that here \cc{X} is a guardian in the guarded read pattern (\eg,
\cc{token} in~\autoref{f:lh-bugs}) and thus it should be persisted
last (after \cc{key} and \cc{value}).

\PP{RA1. Guardian implies persistence atomicity.}
As in the \cc{RO3} ordering invariant, we can find a set of guardians --
\eg, \cc{token[j]} and \cc{token[k]} in~\autoref{f:lh-bugs}. 
%
A program state could be inconsistent if they are not updated atomically.
Based on this observation, we infer the likely invariant on
persistence atomicity such that two or more guardians should be
atomically updated.
\cc{RA1} in \autoref{t:metarule} says: for two guardians \cc{X} and
\cc{Y} from \cc{RO3}, we infer the Likely Invariant \cc{AP(X,Y)} that \cc{X} and
\cc{Y} should be atomically persisted. We simulate NVM states such
that only one guardian is persisted. This approach allows us to reduce
testing space significantly because we will not test persistence
atomicity for well-guarded NVM data. For example, if a program applies
the guarded read patterns on \cc{key} and \cc{value} in all places (using
\cc{token} as a guardian), then we do not test persistence
atomicity between them. 

\begingroup
\addtolength{\tabcolsep}{-3pt}
\renewcommand{\arraystretch}{0.95} 
\begin{table}[t]
  \centering
  \resizebox{\columnwidth}{!}{%
    \begin{tabular}{|c|l|c|c|c|c|c|}
\hline
  \multirow{2}{*}{\textbf{\#}} &
  \multicolumn{2}{c|}{\textbf{Condition}} &
  \multicolumn{2}{c|}{\textbf{Likely Invariant}} &
  \multicolumn{2}{c|}{\textbf{NVM Image}}
\\ \cline{2-7}
  &
  \textbf{Example} &
  \textbf{Rule} &
  \textbf{Example} &
  \textbf{Rule} &
  \textbf{~~~\PST~~~} &
  \textbf{\UPST}
\\ \hline
  \cc{RO1} &
  \cc{Y=X+3;} &
  \cc{W(Y){\dd}R(X)} &
  \cc{X=...;Y=...;} &
  \cc{P(X){\hb}W(Y)} &
  \cc{Y} & \cc{X}
\\ \hline
  \cc{RO2} &
  \cc{\cppX{if}(X)\{Y=3;\}} &
  \cc{W(Y){\cd}R(X)} &
  \cc{X=...;Y=...;} &
  \cc{P(X){\hb}W(Y)} &
  \cc{Y} & \cc{X}
\\ \hline
  \cc{RO3} &
  \cc{\cppX{if}(X)\{Z=Y+3;\}} &
  \cc{R(Y){\cd}R(X)} &
  \cc{Y=...;X=...;} &
  \cc{P(Y){\hb}W(X)} &
  \cc{X} & \cc{Y}
\\ \hline
  \multirow{2}{*}{\cc{RA1}} &
  \cc{\cppX{if}(X)\{M=N+3;\}} &
  \cc{R(N){\cd}R(X)} &
  \multirow{2}{*}{\cc{X=...;Y=...;}} &
  \multirow{2}{*}{\cc{AP(X,Y)}} &
  \cc{X} & \cc{Y}
\\ \cline{6-7}
  &
  \cc{\cppX{if}(Y)\{K=J+3;\}} &
  \cc{R(J){\cd}R(Y)} &
  &
  &
  \cc{Y} & \cc{X}
\\ \hline
\end{tabular}

  }
  \\
  \scriptsize
  \vspace{.5em}
  \cc{R(X)}: read    \cc{X} \quad
  \cc{W(X)}: write   \cc{X} \quad
  \cc{P(X)}: persist \cc{X} \quad
  \PST: persisted           \quad
  \UPST: unpersisted        \quad
  \\
  \cc{E1} \cd \cc{E2}: \cc{E1} is control dependent on \cc{E2} \quad
  \cc{E1} \dd \cc{E2}: \cc{E1} is data dependent on \cc{E2}
  \\
  \cc{E1} \hb \cc{E2}: \cc{E1} should happen before \cc{E2}    \quad
  \cc{AP(X,Y)}: \cc{X} and \cc{Y} persisted atomically
  \vspace{.5em}
  \caption{
    Three meta-rules \cc{RO1}--\cc{RO3} for persistence ordering
    invariants and one meta-rule \cc{RA1} for atomicity invariants.
  }
  \vspace{-10px}
  \label{t:metarule}
\end{table}
\endgroup

\subsubsection{Program Analysis for Invariant Inference}
\label{s:design:invariant:extraction}

\sys performs program dependence analysis to infer likely invariant
from the source codes and execution traces. \sys
first constructs Program Dependence Graph
(PDG)~\cite{pdg-ferrante-toplas87,pdg-ottenstein-sigplan84,pdg-harrold-sigsoft93}
where a node represents a traced instruction, and an edge represents
data or control dependency. Then, \sys simplifies the PDG into what we
called Persistence Program Dependence Graph (PPDG) that captures
dependencies between NVM accesses to make it easy to apply the invariant
inference meta-rules. For example, \autoref{f:design-example}(c) shows
the PDG of the trace (b), and (d) shows the PPDG. 

\sys uses a mix of static and dynamic trace analysis to construct a
PDG. When instrumenting the source code for tracing
(\autoref{s:design:trace}), it performs static analysis to capture
register-level data and control dependency. Then it
extracts memory-level data dependence by analyzing memory-level
data-flow in the collected trace. This dynamic memory-level data
dependency analysis improves PDG's precision compared to static-only
analysis which suffers from the imprecision of pointer analysis. The
static instruction IDs (binary address) are used to map static and
dynamic information.

\sys converts a PDG to a PPDG
as follows. Initially, the PPDG has only (gray) NVM nodes.
\sys traverses the PDG from one NVM node to another NVM node.
If there is at least one control-flow edge along the path, it adds a
control-flow edge in the PPDG. If a path includes only
data-flow edges, it adds a data-flow edge in the PPDG. No path
implies no dependency.

Given the PPDG, \sys then applies the meta-rules in
\autoref{t:metarule} to infer likely invariants. For each edge and two nodes in
the PPDG, \sys considers the type of edge (control vs. data) and the
type of instructions (\store vs. \load). When \sys finds a Condition, it
records the corresponding Likely Invariant.
For example, the PPDG in \autoref{f:design-example}(d) shows that
\cc{R(key)}{\cd}\cc{R(token)}. Based on \cc{RO3}, we infer the invariant
\cc{I1:} \cc{P(key)}{\hb}\cc{W(token)} in (e). Similarly, we can infer the
persistence ordering invariants \cc{I2} and \cc{I3}. Moreover, as
\cc{token} and \cc{key} are guardians for guarded reads, based on
\cc{RA4}, we infer the persistence atomicity invariant
\cc{I4: AP(token,key)}.

\subsection{Generating Crash NVM Images}
\label{s:design:ctp}

The next step after inferring invariants is to generate a set of crash NVM
images\footnote{
  In PMDK, an NVM image is a regular file containing an
  NVM heap state created, loaded, and closed by PMDK APIs~\cite{pmdk-heap}.}
that violate the invariants. Later
in~\autoref{s:design:validation}, we will describe how \sys loads
these NVM images and uses output equivalence checking for validation.


At a high level, \sys generates crash NVM images as follows. \sys
takes as input the same trace used to collect invariants and performs
cache and NVM simulations along the trace. During the simulation,
\sys cross-checks if there is an invariant violation. Each
invariant-violating NVM state forms a crash NVM image. \sys produces
a set of crash NVM images for further validation. 

\subsubsection{Simulating Cache and NVM States}
\label{s:design:ctp:sim}

The goal of the cache/NVM simulations is to generate only feasible NVM
states that violate likely invariants but still obey the semantics of a
persistence control at a cache line granularity (\eg, the effects of a
\flush instruction).
Starting from the empty cache and NVM states, \sys simulates the effects
of \store, \flush, and \fence instructions along the trace while
honoring the memory (consistency) model of a processor. In particular,
\sys supports Intel's x86-64 architecture model, as in
Yat~\cite{yat-lantz-atc14}. The following two rules are, in particular,
relevant to the cache/NVM simulations:

\squishlists
\item A \fence instruction guarantees that all the prior {\flush}-ed
  stores 
  are persisted.
\item A processor does not reorder two store instructions
  in the same cache line (following
  the x86-TSO memory consistency model~\cite{tsox86-sewell-10,intel:sdm}).
  Suppose that \cc{X}
  and \cc{Y} are in the same cache line and a program executes two
  stores \cc{W(X)} and \cc{W(Y)} in order. In this case, the valid NVM
  states are: (1) nothing persisted, (2) only \cc{X} is
  persisted, and (3) both \cc{X} and \cc{Y} are persisted. Note that
  the case that only \cc{Y} is persisted is \emph{not} valid since it
  violates the program order.
\squishends

Consider the trace of Level Hashing's \cc{level_insert} code in
\autoref{f:design-example}(f). After simulating the first three \store
instructions (TID 200-202), there could be multiple valid
cache/NVM states. For example, the data \cc{``k''} for \cc{key} could
either remain in a cache (unpersisted) or could be evicted
(persisted). The same is true for the \cc{val} and \cc{token}.
However, after finishing the execution of the last \fence instruction
(TID 205), \cc{key} and \cc{val} are guaranteed to be persisted (due
to \flush and \fence). Still, \cc{token} could be either unpersisted
or persisted.

\subsubsection{Checking Invariants Violations}
\label{s:design:ctp:chk}

During the simulation, \sys checks if there could be an NVM state that
violates a likely invariant before executing each \fence instruction because the
\fence ensures a persistent state change. \sys considers all possible
persisted/unpersisted states while honoring the above cache/NVM
simulation rules.

Consider the trace of Level Hashing's \cc{level_insert} code in
\autoref{f:design-example}(f) again. Before we execute the last \fence
instruction (TID 205), we check the four invariants against the trace as
shown in (e). For instance, \cc{I1} says that
\cc{P(key)}{\hb}\cc{W(token)}.
The invariant violating state is the one that \cc{token} is
persisted, but \cc{key} is not. We check if this invariant violating case
is feasible in this code region (before the \fence). The answer is yes
-- a program crashes between the TID 202 \store and the TID 203 \flush
instructions, and the cache line for \cc{token} is evicted (persisted)
but not for \cc{key} and \cc{val} (unpersisted). This forms the first
crash NVM image \cc{IMG1} in (g). Similarly, we can find that \cc{IMG1}
is also the state that \cc{I3} and \cc{I4} are violated. We can also
find the second \cc{IMG2} in (g) violating \cc{I2} and \cc{I4}.

Each crash NVM image is indeed represented as a pair
of a fence ID and a store ID, which specifies where to crash and
which store to be persisted, respectively. \sys repeats the process along
the trace and generates a set of crash NVM images that will be
validated in the next step.

\subsection{Output Equivalence Checking}
\label{s:design:validation}

\sys validates the invariant-violating crash NVM images and detects crash
consistency bugs using output equivalence checking.
%
The key idea is that if
the NVM program is crash-consistent, after a recovery from a crash, it
should behave as if the operation where the crash occurred is either
fully executed (committed) or not at all executed (rollbacked). And
thus, after the crash, the program should produce the output of one of
these two executions, which we call \emph{oracles}. 

Consider the example in \autoref{f:design-example} again. Using the
test case \cc{insert(k,v0)}, \cc{delete(k,v0)}, \cc{insert(k,v1)}, and
\cc{query(k)} in (a), we analyzed the trace of the third
\cc{insert(k,v1)} operation in (f) to generate two crash NVM images in
(g). The first \cc{IMG1} reflects an NVM state that the first two
operations, \cc{insert(k,v0)} and \cc{delete(k,v0)}, are correctly
performed, and the program crashes in the middle of the third
\cc{insert(k,v1)} where only \cc{token} is persisted, and \cc{key} and
\cc{value} remain unpersisted -- \ie, \cc{IMG1} has the old value \cc{v0}.

\sys generates two oracles to compare. The first oracle
reflects an execution where the crashed operation is committed -- thus
we run the test case \cc{insert(k,v0)}, \cc{delete(k,v0)},
\cc{insert(k,v1)}, and \cc{query(k)} (no crash) and records \cc{v1}
(the new value) as the output of \cc{query(k)}. The second oracle
mimics an execution where the crashed operation is rollbacked
-- we run the same test case without the third \cc{insert(k,v1)} and log
\cc{null} as the output of \cc{query(k)}. Altogether, the oracles say
that the correct output of the last \cc{query(k)} is either \cc{v1} or
\cc{null}.

For output equivalence checking, \sys loads a crash NVM image, runs a
recovery code (if exists), executes the rest of the test cases,
records their outputs, and compares them with the oracles. For example
with \cc{IMG1}, \cc{query(k)} returns the old value \cc{v0} (as
neither the deletion of \cc{k} nor the insertion of new value \cc{v1}
was not persisted) -- \sys detects the mismatch and 
reports the test case and the crash NVM image information (the crash
location as the \fence TID, and the persistence state as the persisted
\store ID).
On the other hand, a similar analysis with the second \cc{IMG2} shows
that the output (\cc{null}) matches the oracles, so \sys does not
report them.

\subsection{Clustering Bug Reports}
\label{s:design:clustering}


One key benefit of the prior output equivalence checking is that all
the reported cases indeed reflect buggy inconsistent states (no false
positives). Nonetheless, 
many cases may share the same
root cause: \eg, a bug in \cc{insert} operation may repeatedly appear
if the test case has many \cc{insert} calls.

To help programmers finding the root causes, \sys clusters the bug
reports according to operation type (\eg, \cc{insert}, \cc{delete})
and execution path (a sequence of basic blocks) that appeared in the
trace.
We found that our clustering scheme helps
the root cause analysis significantly because after one root
cause is found, reasoning about the redundant cases along the same
program path is relatively simpler.
Multiple clusters may share the same root cause.

\subsection{Extension for Multi-threaded NVM Programs}
\label{s:design:multithread}


Comparing two oracles is sufficient for testing single-threaded NVM program.
However, for multithreaded cases, output equivalence checking should consider
more oracles as a program may crash while \cc{M} concurrent operations are
running. Each per-thread operation has two legal states (all or nothing),
and we also need to consider different permutations of a linearization order.
Thus, the number of oracles is $P(M,0)
+ P(M,1) + ... + P(M,M-1) + P(M,M)$, where $P(m,k) =
\sfrac{m!}{(m-k)!}$. For example, we need 5 oracles for 2 threads and
65 oracles for 4 threads.
As discussed in~\autoref{s:design:tc}, \sys focuses on exploring the
persistence space in a systematic and scalable manner, and we leave
thread-interleaving space reduction as future work.

\begin{table*}[t]
  \centering
  \resizebox{\textwidth}{!}{\begin{tabular}{|c||c||c|c|c|c||c|c|c||c|c|c|c||c|c|}
\hline

&
\multirow{2}{*}[-12pt]{\textbf{Application}} &
\multirow{2}{*}[-12pt]{\textbf{Description}} &
\multirow{2}{*}[-12pt]{\textbf{Design}} &
\multirow{2}{*}[-12pt]{\textbf{\makecell{Concu- \\ rrency}}} &
\multirow{2}{*}[-12pt]{\textbf{LOC}} &
\multicolumn{3}{c||}{\textbf{Likly Invariant Inference}}&
\multicolumn{4}{c||}{\textbf{Output Equivalence Checking}} &
\multicolumn{2}{c|}{\textbf{\# bugs}} \\ \cline{7-15}
&
&
&
&
&
& \textbf{\makecell{\# ordering \\ invariants}}
& \textbf{\makecell{\# atomicity \\ invariants}}
& \textbf{\makecell{execution \\ time}}
& \textbf{\makecell{\# crash \\ NVM \\ images}}
& \textbf{\makecell{\# images w/ \\ output \\ mismatch}}
& \textbf{\makecell{\# \\ clusters}}
& \textbf{\makecell{execution \\ time}}
& \textbf{\makecell{\# \\ PO}}
& \textbf{\makecell{\# \\ PA}}
\\ \hline \hline

\multirow{10}{*}{\rotatebox[origin=c]{90}{\textbf{Persistent Data Structures}}}
& WOART~\cite{wort-lee-fast17}                  & radix tree          & LL-ASM  & ST & 835   & 33428  & 5027  & 1m51s  & 13364  & 22       & 5   & 2m33s   & 0    & 1   \\ \cline{2-15}
& WORT~\cite{wort-lee-fast17}                   & radix tree          & LL-ASM  & ST & 498   & 21248  & 4709  & 53s    & 18886  & 0        & 0   & 2m42s   & 0    & 0   \\ \cline{2-15}
& Fast Fair~\cite{fast-fair-hwang-fast18}       & B+tree              & LL-ASM  & LB & 1083  & 468865 & 1318  & 3m58s  & 59270  & 46866    & 104 & 12m21s  & 1    & 2(1)\\ \cline{2-15}
& Level Hash~\cite{level-hashing-zuo-osdi18}    & hash table          & LL-ASM  & ST & 1008  & 27941  & 1653  & 6m41s  & 52678  & 43177    & 29  & 1h40m   & 10   & 7   \\ \cline{2-15}
& CCEH~\cite{cceh-nam-fast19}                   & hash table          & LL-ASM  & LB & 989   & 8960   & 1842  & 3m17s  & 18697  & 490      & 5   & 30m3s   & 0    & 2(1)\\ \cline{2-15}
& P-ART~\cite{recipe-lee-sosp19}                & radix tree          & LL-ASM  & LB & 2356  & 2065   & 1389  & 2h     & 16454  & 36       & 4   & 3m27s   & 0    & 2   \\ \cline{2-15}
& P-BwTree~\cite{recipe-lee-sosp19}             & B+tree              & LL-ASM  & LF & 4860  & 35135  & 5400  & 47m53s & 41662  & 5137     & 85  & 1h11m   & 2    & 0   \\ \cline{2-15}
& P-CLHT~\cite{recipe-lee-sosp19}               & hash table          & LL-ASM  & LB & 4461  & 3065   & 806   & 2m2s   & 6086   & 2        & 1   & 9m33s   & 1    & 0   \\ \cline{2-15}
& P-HOT~\cite{recipe-lee-sosp19}                & trie                & LL-ASM  & LB & 10463 & 31435  & 18608 & 4h     & 72998  & 1253     & 111 & 25m26s  & 3    & 0   \\ \cline{2-15}
& P-Masstree~\cite{recipe-lee-sosp19}           & B tree + trie       & LL-ASM  & LB & 1975  & 22662  & 4119  & 10m33s & 75234  & 76       & 10  & 12m     & 0    & 1   \\ \cline{2-15} \hline \hline
\multirow{5}{*}{\rotatebox[origin=c]{90}{\textbf{Micro}}}
& Array                                         & array               & LL-PMDK & ST & 405   & 3208   & 488   & 6m35s  & 488    & 487      & 2   & 30s     & 0    & 1(1)\\ \cline{2-15}
& B-Tree                                        & B tree              & TX-PMDK & ST & 493   & 1143   & 130   & 28m46s & 97590  & 21551    & 50  & 20m57s  & 1*   & 1(1)\\ \cline{2-15}
& C-Tree                                        & crit-bit trie       & TX-PMDK & ST & 293   & 9752   & 704   & 58m49s & 23875  & 0        & 0   & 16m31s  & 0    & 0   \\ \cline{2-15}
& RB-Tree                                       & red-black tree      & TX-PMDK & ST & 416   & 15337  & 725   & 37m15s & 256470 & 4785     & 64  & 46m16s  & 0    & 1(1)\\ \cline{2-15}
& Hashmap-TX                                    & hash table          & TX-PMDK & ST & 353   & 8979   & 801   & 1h     & 25997  & 3        & 2   & 27m4s   & 1    & 0   \\ \hline \hline
\multirow{2}{*}{\rotatebox[origin=c]{90}{\textbf{Real}}}
& Memcached                                     & key-val store       & LL-PMDK & LB & 17731 & 11228  & 2708  & 14m40s & 11493  & 0        & 0   & 48h28m  & 0    & 0   \\ \cline{2-15}
& Redis                                         & key-val store       & TX-PMDK & ST & 44642 & 5659   & 942   & 1h35m  & 174610 & 0        & 0   & 72h45m  & 0    & 0   \\ \hline \hline
& \textbf{Total}                                & -                   & -       & -  & 92861 & 710110 & 51369 & 12h19m & 965852 & 123885   & 472 & 127h34m & 19   & 18(5)\\ \hline
\end{tabular}
}

  \vspace{.5em}
  \scriptsize
  \textbf{LL-ASM}: low-level persistence primitives (asm) ~~~
  \textbf{TX-PMDK}: PMDK transaction ~~~
  \textbf{LL-PMDK}: low-level persistence primitives (PMDK)

  \textbf{ST}: single-threaded ~~~
  \textbf{LB}: lock-based ~~~
  \textbf{LF}: lock-free ~~~
  \textbf{PO}: persistence ordering bug ~~~
  \textbf{PA}: persistence atomicity bug ~~
  \textbf{(\#)}: number of known bugs ~~~
  \textbf{*}: bug in PMDK library
  \vspace{6pt}
  \caption{
    The tested NVM programs, the detailed statistics of \sys bug finding, and the number of detected bugs.
  }
  \vspace{-8px}
  \label{t:bugstats}
\end{table*}

\section{Implementation}
\label{s:impl}


All \sys components except tracing and program dependency analysis are
written in 4400 lines of Python code.
We built tracing and program dependency analysis based on
Giri~\cite{giri-sahoo-asplos13}, a dynamic program slicing tool
implemented in LLVM~\cite{llvm-web}. Our Giri
modification comprises of around 3600 lines of C++ code and includes
the following three extensions. First, we ported Giri from LLVM v3.4 to
v9.0.1. Second, we extended its program slicing component to generate
PDG and PPDG. Third, we modeled library function calls and assembly
instructions that have the semantics of load/store (\eg,
\cc{atomic.store}) and persistence (\eg, \cc{clwb}, \cc{sfence}).

Our current prototype supports an NVM program built on PMDK
\cc{libpmem} or \cc{libpmemobj} libraries to create/load an NVM image
from/to disk. To ensure the virtual address of \cc{mmap}-ed NVM heap
the same across different executions, we set \cc{PMEM_MMAP_HINT}
environment variable. \sys runs PPDG construction, crashed NVM image
generation, and output equivalence checking in parallel.


The current prototype does not support kernel-level NVM
programs. Tracing a kernel execution and
inferring invariants can be supported. More engineering
efforts, however, are required to support checkpointing (or rollback) kernel/NVM
states and resuming from a clean state for output equivalence
checking. 
Systematic testing of a kernel-level
program (\eg, using virtualization like Yat) is left as future work.

\section{Evaluation}
\label{s:eval}

%


\subsection{Evaluation Methodology}
\label{s:eval:method}

%

\PP{NVM Programs.}
\autoref{t:bugstats} shows the three groups of 17 NVM programs
that we used to evaluate \sys.
The first group includes ten state-of-the-art persistent data
structures highly optimized for NVM and published at top-tier systems
conferences. 
For high performance,
they all used \emph{low-level} (LL) persistence primitives such as
\flush and \fence instructions.
Some (\eg,
FAST-FAIR~\cite{fast-fair-hwang-fast18}) incorporate inconsistency
tolerable design where a naive crash consistency bug detection
approach would lead to false positives. 

We found that they are not properly written to
use persistent heap and locks for NVM. They were all \emph{emulated
with volatile memory}. Thus, we made the following source code changes
to make them actually use NVM. 
First, they all used either regular \emph{volatile} memory allocator (\ie,
\cc{malloc}, \cc{free}) or PMDK's \cc{libvmmalloc}
allocator~\cite{pmdk-vmmalloc}, which
do not guarantee crash consistency of the NVM heap metadata.
To faithfully construct an NVM image including the
NVM heap state, we ported them to use PMDK's \cc{libpmemobj}
allocator~\cite{pmdk-alloc}.
Second, for lock-based data structures, we added a recovery
code to release locks when loaded from an NVM image to avoid 
deadlock.
%
%
We made 5041 lines of code and script changes in total.
With the advent of real NVM hardware (\eg, Intel’s Optane memory), we
believe all reasonable (future) applications should use a proper
persistence library (\eg, PMDK).
For applications that are already implemented with NVM/PMDK in the
following groups, we did not make any change.

The second group includes four relatively-simple (300--500 LOC)
persistent data structures that appeared in the PMDK library as
example codes. They used PMDK's low-level (LL) or \emph{transactional} (TX)
persistence programming model. We included them mainly to 
compare \sys with prior works which only test them and do not evaluate the first group.
For TX-PMDK applications, \sys traces and analyzes PMDK internals such as
persistence heap allocation and transactional undo logging logics, validating
both the PMDK library itself and transactional NVM programs using them.

The last group includes PMDK-based two server programs
\cc{Memcached} and \cc{Redis} using PMDK's LL and TX persistence APIs,
respectively.

\PP{Test case}
We run the NVM programs with a test case consisting of 2000
randomly generated operations. They provide a different
but mostly similar set of APIs (\eg, \cc{insert}, \cc{delete},
\cc{query}).
The server programs were also tested with the client generating the
same number (2000) of random key-value requests over network.
As discussed in~\autoref{s:design:tc}, \sys does not focus on exploring program
input test space, and simply relies on a random testing.
%
We found that 2000 operations are
large enough to achieve a reasonable and stable code coverage
(50\%-80\%). Missing code coverages are due to unused features (\eg,
garbage collection) and debugging codes. 
%

\PP{Experimental setup}
We ran all experiments on a 64-bit Fedora 29 machine with two 16-core
Intel Xeon Gold 5218 processors (2.30GHz), 192 GB DRAM, and 512 GB
NVM.

\subsection{Detected Crash Consistency Bugs}
\label{s:eval:newbugs}

%

\sys detected  \NumOfTotalBugs (\NumOfNewBugs new) bugs from \NumOfAppsBugDetected programs. 
There were \NumOfOrderBugs
persistence ordering bugs and \NumOfAtomicityBugs persistence atomicity bugs. 
All the bugs were confirmed by the developers.
See the last two columns of \autoref{t:bugstats}.
%

The detected bugs have diverse impacts: lost, unexpected, duplicated
key-value pairs; unexpected operation failure; and inconsistent
structure.
%
%
For example, a crash in the middle of rehashing operation in Level
Hashing 
may lead to lost, unexpected, duplicated key-value pairs since
the metadata is not consistent with the stored key-value pairs.
In FAST-FAIR,
if a crash happens in splitting the root node
and right before setting the new root node, the B+tree will be in an
illegal state; the root node connects to a sibling node. Any further
operation on the B+tree will lead to a program crash or performance degradation.

Many detected bugs are not shallow. For instance, the bug in CLHT 
only occurs when a program crashes at a
specific moment during rehashing while leaving a specific set of
stores unpersisted. Our study reveals that it is hard for a
developer to reason about all possible NVM states (as reasoning about
all possible thread interleaving is difficult for multithreaded
programming).

The bug appeared in B-Tree 
was indeed a persistence ordering bug
inside of PMDK's persistent pool/heap allocation function
\cc{pmemobj\_tx\_zalloc}~\cite{btree-issue-0}. The bug did not manifest in
other TX-PMDK applications as it resides in a code path that requires a
large-size object allocation.

\PP{Comparison with RECIPE}
Our results show that \sys's approach is effective in discovering new
crash consistency bugs in NVM programs. In particular, all the tested low-level
persistent data structures (except for WOART and WORT) have been heavily
tested by RECIPE~\cite{recipe-lee-sosp19}, but \sys is still able to
report 30 new bugs.
RECIPE only reported four bugs, two of which 
are
overlapped with \sys. The other two bugs are due to missing
persistence primitives in root node initialization. We found and fixed
these two bugs while we ported the test data structures to use PMDK's
memory allocator, so we did not count them as the bugs that \sys found.

\subsection{Statistics of \sys Bug Finding}
\label{s:eval:stat}


\autoref{t:bugstats} also presents the detailed statistics of each major step
in \sys.
%
Across \NumOfAppsTested NVM programs, when tested with 2,000 operations, 
\sys infers in total 710K (42K on average) ordering invariants and 51K (3K) atomicity invariants. 
\sys generated 966K (57K) crash NVM images, 124K of which failed output equivalence checking.

\sys finally reported 472 clustered bugs.
We found that bug clustering
significantly reduces manual efforts in root cause
analysis.
\sys also provides sufficient information for root cause analysis
including execution trace, crash location, persisted and unpersisted
writes, and a crash NVM image which can be loaded for further \cc{gdb}
debugging.
Two graduate students performed root cause analysis.
It took about 7 hours to investigate the 472 bug clusters.

We found that \sys prototype is fast enough for practical use. Invariant
inference took a few minutes to four hours. Output equivalence checking took a
few minutes to two hours, except for two servers. 
The overhead of output equivalence checking is proportional to the number of
crash NVM images and the cost of each test run.
As server programs, \cc{Memcached} and \cc{Redis} require live
networking-based testing, which made output equivalence checking
slower than the others. Its high overhead stems from server
start/shutdown and client connection setup cost for each test run. The
pure execution overhead was indeed small as in other
applications. Moreover, the current prototype does not parallelize
output equivalence checking for server programs.

\subsection{Comparison with Annotation-Based Approaches}
\label{s:eval:annotation}

\begingroup
\addtolength{\tabcolsep}{-3pt}
\renewcommand{\arraystretch}{0.92} 
\begin{table}[t]
  \centering
  \resizebox{\columnwidth}{!}{

\begin{tabular}{l|c|c|c|c|l}
\hline
      & \multirow{2}{*}{\textbf{App}} & \multirow{2}{*}{\textbf{Design}} & \multicolumn{2}{c|}{\textbf{Bugs detected by}} & \multirow{2}{*}{\textbf{Bug Description}} \\ \cline{4-5}
      & & & \textbf{\sys?} & \textbf{Others?} & \\ \hline \hline
\multirow{4}{*}{\rotatebox[origin=c]{90}{\textbf{Micro}}}
      & B-Tree     & TX-PMDK & \yesmark & \yesmark                   & Modify without logging \\
      & C-Tree     & TX-PMDK & \nomark  & \nomark                    & N/A                    \\
      & RB-Tree    & TX-PMDK & \yesmark & \yesmark                   & Modify without logging \\
      & Hashmap-TX & TX-PMDK & \nomark  & \nomark                    & N/A                    \\ \hline \hline
\multirow{2}{*}{\rotatebox[origin=c]{90}{\textbf{Real}}}
      & Memcached  & LL-PMDK & \nomark  & \nomark                    & N/A                    \\
      & Redis      & TX-PMDK & \nomark  & \yesmark \textbf{(Benign)} & Modify outside TX      \\ \hline
\end{tabular}
}

  \vspace{.5em}
  \scriptsize
  TX-PMDK: PMDK transaction ~~
  LL-PMDK: PMDK low-level persistence primitives
  \vspace{6pt}
  \caption{
    Comparison between \sys and two annotation-based approaches,
    PMTest~\cite{pmtest-liu-asplos19} and
    XFDetector~\cite{xfdetector-liu-asplos20}.
  }
  \vspace{-12px}
\label{t:annotation}
\end{table}
\endgroup

This section compares \sys with two
annotation-based approaches, PMTest~\cite{pmtest-liu-asplos19} and
XFDetector~\cite{xfdetector-liu-asplos20}.
%
%
%
PMTest provides two primitives:
\cc{isPersist} and \cc{isOrderedBefore}, but does not support one for
persistence atomicity. Hypothetically, assuming correct/full annotations,
PMTest would be able to detect \NumOfOrderBugs ordering bugs but miss
\NumOfAtomicityBugs atomicity bugs in \autoref{t:bugstats}.
XFDetector targets logging-based (\eg, UNDO,
REDO) NVM programs. XFDetector cannot be applied to the 12 low-level programs.
Besides, a developer should annotate ``commit variable'' to prune benign cases.

Actually, 6 of \NumOfAppsTested programs in \autoref{t:bugstats} were tested
by PMTest and XFDetector as well.
\autoref{t:annotation} lists those six programs that consist of four
micro-benchmarks from PMDK examples and two servers (\cc{Redis} and
\cc{Memcached}). 
%
More precisely, PMTest and XFDetector tested one and two more
programs: HashMap-Atomic and PMFS. We did not test HashMap-Atomic
because it relies on PMDK's persistent linked-list library, which we
did not instrument and trace. We also did not test PMFS, an in-kernel
file system, because \sys currently only supports user-space
applications and the PMFS code~\cite{pmfs-github} has not been
maintained for real NVM.
We expect that \sys should be able to detect bugs in HashMap-Atomic and PMFS
by extending the prototype because they share the same
characteristics as the bugs detected in other tested programs.
%
%

\PP{Detected bugs}
\autoref{t:annotation} shows that among the three bugs that
PMTest/XFDetector found, \sys also detects two of them in B-Tree 
and RB-Tree.
Both are due to missing logging inside a transaction.
Note that \sys detected another bug
in the PDMK library while
testing B-Tree, which was missed by them.

\PP{Missed bug}
\sys missed one bug in \cc{Redis} reported by PMTest/XFDetector. In theory,
\sys may have false negatives for two reasons: (1) \sys may not be
able to infer the relevant likely invariants from the trace. (2) The test case
may not reveal the symptom of inconsistent behavior during output
equivalence checking.
Upon further investigation, we found that it is none of the two
cases. Interestingly, it turns out that the bug was benign.
The bug is in the server initialization code. After allocating a PMDK
root object, \cc{Redis} initializes the root object to zero ``outside'' of
a PMDK transaction. PMTest/XFDetector detects this unprotected update
as a bug. However, this is benign -- it does not lead to an
inconsistent state. The root object was allocated
using \cc{POBJ_ROOT()}~\cite{pobjroot-web}, which already zeroed out
the newly allocated object. Both the old and new values are
zero. Therefore, it does not matter if the new zero update is
persisted or not.
\sys actually detected this store violating an atomicity invariant, and
performed output equivalence checking. But it does not show any visible
divergence.
This example particularly shows the benefit of our output
equivalence checking, pruning false positives.

\subsection{Comparison with Exhaustive Testing Approaches}
\label{s:eval:ts}

\begin{figure}
  \centering
  \vspace{3px}
  \input{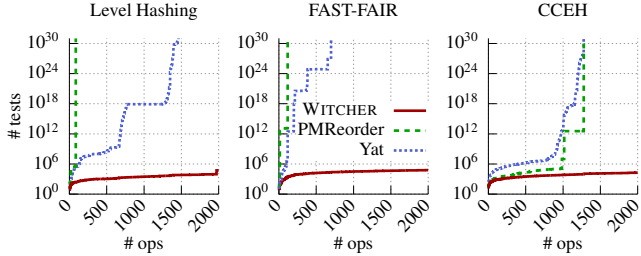}
  \vspace{-8px}
  \caption{
    Test space comparison for 2000 random operations.
  }
  \vspace{-10px}
  \label{f:ts}
\end{figure}

Lastly, we compare \sys with two existing exhaustive-testing-based
tools Yat~\cite{yat-lantz-atc14} and
PMReorder~\cite{pmreorder-intel-web} to show how effectively our
likely invariant-based approach can reduce the testing space.
In particular, we compare the number of crash states that each tool
will validate using the same trace with 2000 random operations as
in~\autoref{s:eval:newbugs}. We simulate Yat and PMReorder algorithms
to calculate the numbers of test cases.

\autoref{f:ts} shows the representative results for Level Hashing,
FAST-FAIR, and CCEH programs. The test space of Yat and PMReorder is
several orders larger than \sys. Sudden spikes happen in Yat when
there is a rehashing in Level Hashing and CCEH or a node
split/merge in FAST-FAIR. That is because rehashing and node
split/merge require many key-value movements, and Yat tests all
possible crash states due to its exhaustive approach. On the other
hand, \sys only tests when there is an invariant violation, significantly
reducing the number of test cases (yet detecting many bugs).

PMReorder behaves much worse than Yat particularly in FAST-FAIR and
Level Hashing because it does not perform cache line
granularity analysis. As a result, PMReorder may test many infeasible
crash states violating the memory model, hinting potential false
positives. PMReorder behaves better than Yat in CCEH before 1278
operations because it only tests those stores that are
explicitly flushed at each fence instruction, indicating potential
false negatives.

\section{Related Work}
\label{s:rlewk}


\PP{Likely-invariants based testing}
A concept of likely-invariants has been used to detect program
bugs~\cite{deviant-behavior-engler-sosp01, afg-kremenek-osdi06, muvi-lu-sosp07,
apisan-yun-sec16, mining-khairunnesa-oopsla17, juxta-min-sosp15}, to verify the
network~\cite{non-lopes-nsdi15}, and to detect resource
leak~\cite{tracker-torlak-icse2010}.
Notably, Engler \etal's version (called
beliefs)~\cite{deviant-behavior-engler-sosp01} enables automatic analysis of
likely correctness conditions without in-depth knowledge.
To the best of our knowledge, \sys is the first work that infers likely
invariant in the context of crash consistency testing for NVM programs.

\PP{Output-equivalence based testing}
Burckhardt \etal~\cite{line-up-burckhardt-pldi10} and Pradel
\etal~\cite{cut-pradel-pldi12}
detect thread-safety violations, comparing the concurrent execution to
linearizable executions of a test.
\sys's output equivalence checking shares some idea of these two works in the
sense that they all compare an observed execution with ``oracles'', but is
uniquely designed to detect for NVM crash consistency bugs.
%

\PP{Crash consistency testing in file systems}
There has been a long line of research in testing and guaranteeing
crash consistency in file systems~\cite{b3-mohan-osdi18, yang:fisc,
  yang:explode, gunawi:eio, rubio:wpds, fscq-chen-sosp15,
  dfscq-chen-sosp17, yggdrasil-sigurbjarnarson-osdi16, janus-xu-sp19,
  hydra-kim-sosp19}.
In-situ model checking approaches such as EXPLODE~\cite{yang:explode}
and FiSC~\cite{yang:fisc} systematically test every legal action of a
file system with minimal modification.
B3~\cite{b3-mohan-osdi18} performs exhaustive testing within a bounded
space, which is heuristically decided based on
the bug study of real file systems.
\sys reduces test space by using inferred invariants, unlike limiting
testing space
in B3.
Feedback-driven File system fuzzers, such as
Janus~\cite{janus-xu-sp19} and Hydra~\cite{hydra-kim-sosp19}, mutate
both disk images and file operations to thoroughly explore file system
states.
We believe \sys's test case generation can be further improved by
adopting feedback-driven fuzzing techniques.

\section{Conclusion}
\label{s:conclusion}


We present \sys, a scalable, automatic, and precise crash consistency
bug detector for NVM software. \sys infers \emph{likely invariants} that are
believed to be true to be crash-consistent from source codes and program
traces, and performs \emph{output equivalence checking} to validate likely
invariant violations.
This approach allows \sys to detect crash consistency bugs without manual
annotations, user-provided consistency checker, or exhaustive testing.
We evaluated \sys on \NumOfAppsTested NVM programs and found
\NumOfTotalBugs crash consistency bugs, including \NumOfNewBugs new ones. 
We will open-source \sys for NVM programmers to use and extend it.

%


\balance
\bibliographystyle{plainurl}
\bibliography{p,manycore,sys,nvm,sec,bugs,tips,witcher,conf}


\end{document}